\documentclass[aip,showpacs,onecolumn,11pt]{revtex4}%
\usepackage{amsmath}
\usepackage{amsfonts}
\usepackage{amssymb}
\usepackage{graphics}
\usepackage{graphicx}
\providecommand{\U}[1]{\protect\rule{.1in}{.1in}}

\begin{document}
\title{Full coherent frequency conversion between two microwave propagating modes}
\author{Baleegh Abdo}
\email{baleegh.abdo@yale.edu}
\author{Katrina Sliwa}
\author{Flavius Schackert}
\author{Nicolas Bergeal}
\altaffiliation{Current address: LPEM-UMR8213/CNRS-ESPCI ParisTech-UPMC, 10 rue Vauquelin-75005 Paris, France.}
\author{Michael Hatridge}
\author{Luigi Frunzio}
\author{A. Douglas Stone}
\author{Michel Devoret}
\affiliation{Department of Applied Physics, Yale University, New Haven, CT 06520, USA.}
\date{\today }

\begin{abstract}

We demonstrate full frequency conversion in the microwave domain using a Josephson three-wave mixing device pumped at the difference between the frequencies of its fundamental eigenmodes. By measuring the signal output as a function of the intensity and phase of the three
input signal, idler and pump tones, we show that the device functions as a controllable three-wave beam-splitter/combiner for propagating microwave modes, in accordance with theory. Losses at the full conversion point are found to be less than $10^{-2}$. Potential applications of the device include quantum information transduction and realization of an ultra-sensitive interferometer with controllable feedback.

\end{abstract}

\maketitle

\newpage

A quantum information transducer capable of converting the frequency of a quantum signal without introducing noise is one of the desirable modules in quantum communication \cite{Painter,OptoToMwLehnert}. With such a device, one could teleport quantum superpositions of ground and excited states of qubits from one system to another one with a different transition frequency, without loss of coherence. If a quantum signal can be routed through different frequency channels, optimization of quantum calculation and communication can be advantageously separated. 

The simplest scheme for performing frequency conversion is pumping a dispersive medium with nonlinearity $\chi_{2}$ at precisely the frequency difference between the input and output frequencies \cite{ObservQFC,VisibleZaske,Ates,QuantumRakher,PolarizationRamelow,QSJAumentado,QIJAumentado,Sipe}. Such nonlinear process known as parametric frequency conversion holds the promise of converting the frequency in a unitary, therefore noiseless, manner, namely, without adding loss or dephasing to the processed signal \cite{TuckerPEC}. In practice however, conversion in the optical domain can be unitary only in the limit of very small converted portion of the input signal. This is due to conversion losses associated with the nonlinear wave mixing process, such as finite interaction lengths and imperfect phase-matching. Similarly, resistive elements used in mixers in the microwave domain, such as Schottky diodes inevitably lead to inacceptable conversion losses. This raises the question as to whether a unitary full conversion of an input signal can be realized practically. 
  
In this work, we show that such noiseless full conversion is possible in the microwave domain by operating a dissipationless three-wave mixing element known as the Josephson parametric converter (JPC) \cite{JPCnature,Jamp,JRCAreview}, in a regime of conversion without photon gain. The JPC performs frequency conversion from $8$ to $15$ $\operatorname{GHz}$ with losses below 0.05 dB at full conversion, as opposed to a typical loss of $6$ dB in microwave mixers. We reveal the unitary nature of the JPC conversion by utilizing wave interference between the three incommensurate frequencies intervening the device. The coherence of the conversion process is thus verified without having to face the challenge of calibrating the transmission between different input and output lines of disparate frequencies.


\begin{figure*}
[htb]
\begin{center}
\includegraphics[
width=0.65\textwidth
]%
{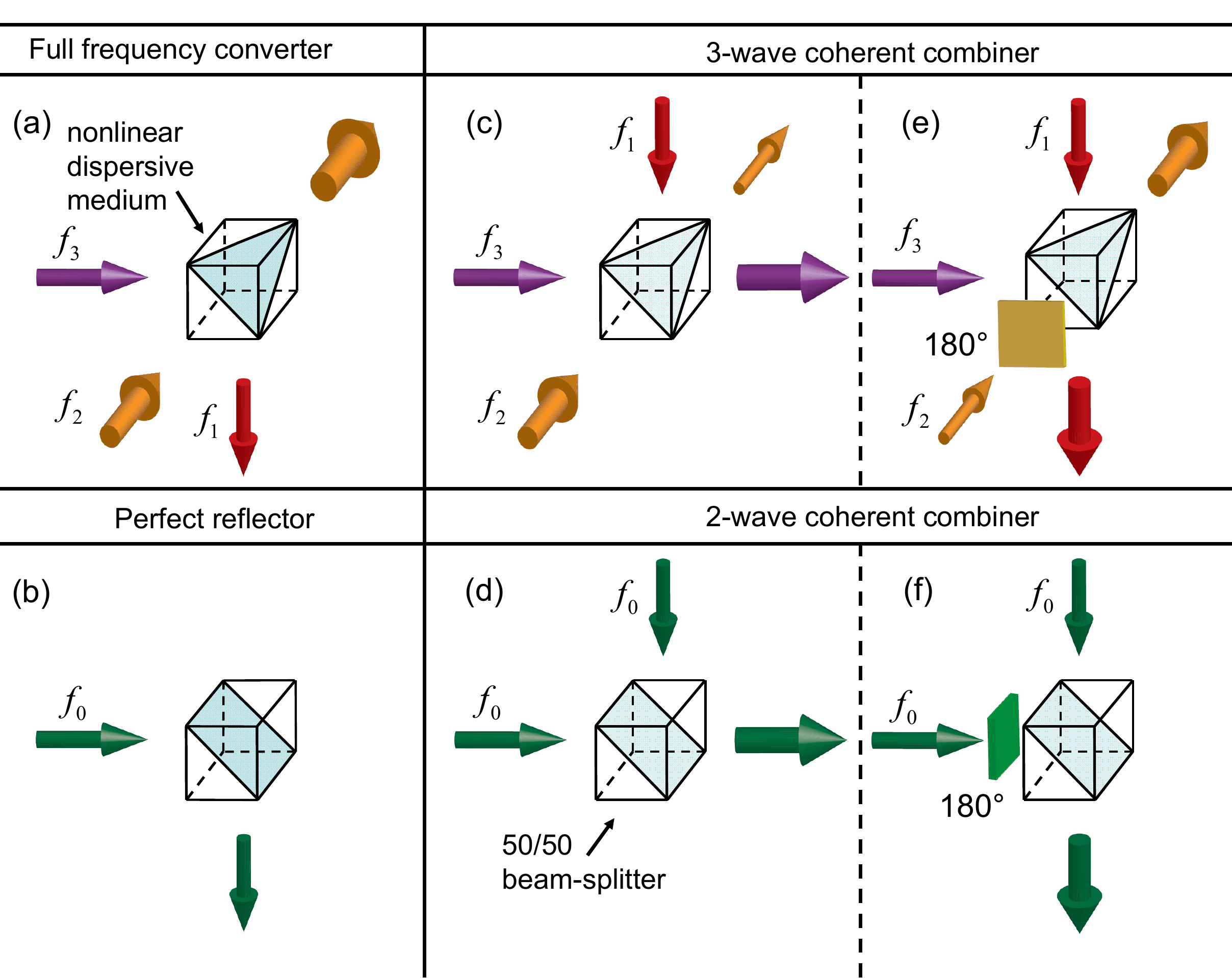}%
\caption{(color online). In a 3-wave coherent converter, the scattered
beams have different frequencies which we indicate using different colors: violet, red and yellow. Panel (a) depicts a nonlinear $\chi_{2}$ dispersive medium operated as a full frequency converter. An incident violet beam is entirely converted into an output red beam via energy exchange with the incident yellow beam playing here the role of the pump. Panel (b) depicts a splitter modified into a perfect mirror which constitutes the 2-wave analogue of the full frequency converter. Panels (c) and (e) introduce the relationship between frequency conversion and 3-wave interference. By operating the device, shown in panel (a), as a 3-wave $50/50$ beam-combiner, one can coherently interfere two beams (red and violet) with disparate frequencies via frequency conversion through the third beam (yellow). Whether a violet beam (panel (c)) or a red beam (panel (e)) is
generated at the output depends on the precise setting of all three relative phases of the incident beams.
Conversely, in a 2-wave coherent beam-splitter/combiner, interference takes place (panels (d) and (f)) when two equal intensity beams with the same frequency are incident at the beam-splitter. The intensity of the outgoing beams depend on the phase difference between the incoming waves. 
}%
\label{BS}%
\end{center}
\end{figure*}


\begin{figure}
[htb]
\begin{center}
\includegraphics[
width=0.49\textwidth
]%
{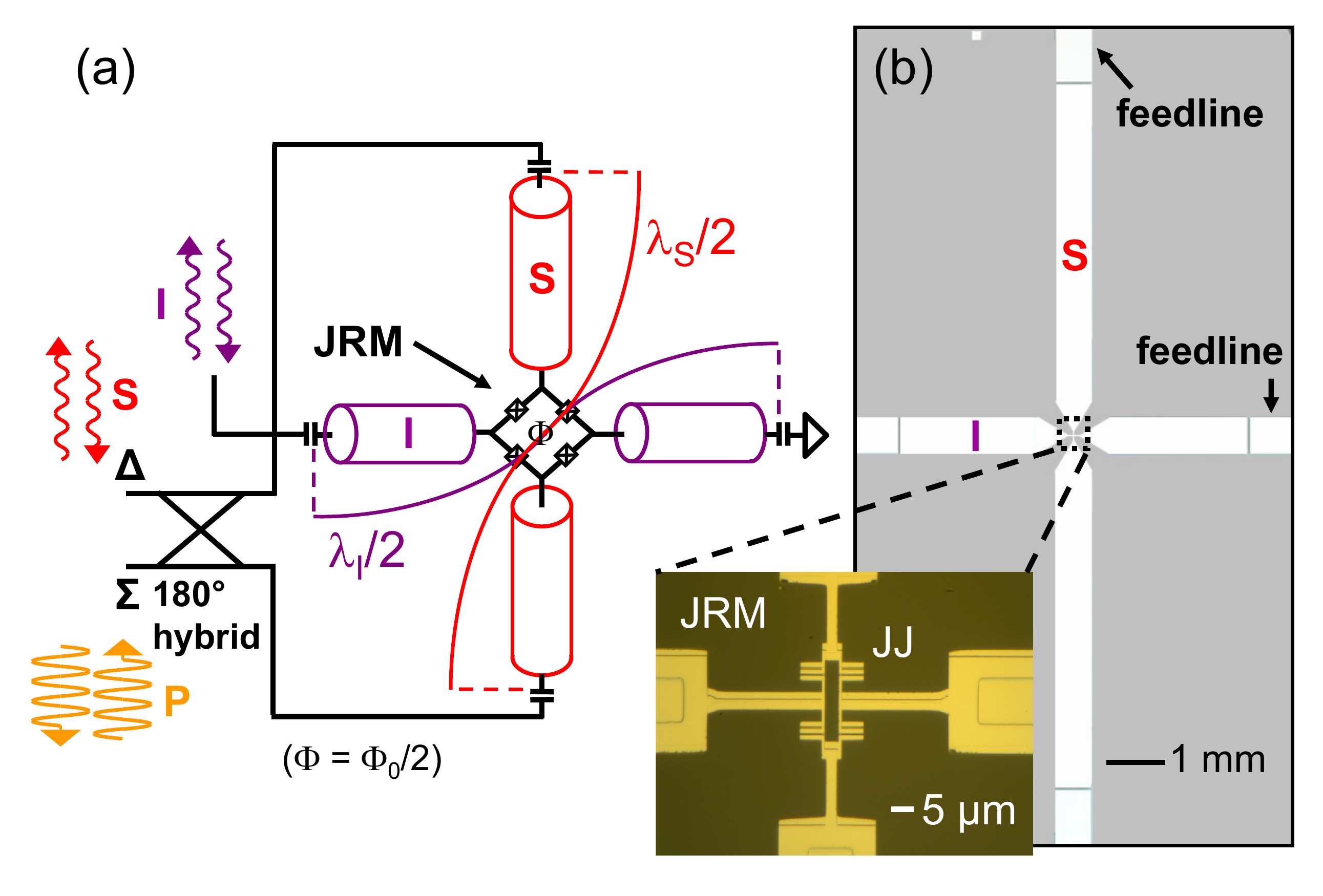}%
\caption{(color online). Panel (a): circuit diagram of the JPC device. The JPC consists of
two half-wave microstrip resonators denoted (S) and (I) which resonate at
$f_{S}=8.402\operatorname{GHz}$ and $f_{I}=14.687\operatorname{GHz}$ with
bandwidths of $95$ $\operatorname{MHz}$ and $270$ $\operatorname{MHz}$
respectively. These are set by the coupling capacitors between the resonators
and the feedlines. The two resonators intersect at a Josephson ring modulator
(JRM), which consists of four nominally identical Josephson junctions arranged in Wheatstone bridge configuration. The JRM is biased near half a flux quantum. The S, I and P coherent waves are represented in this schematic using the same color
scheme as Fig. 1 although in this device, $f_{P}<f_{S}$. Panel (b):
false color optical micrographs of the JPC device and the JRM.}%
\label{setup}%
\end{center}
\end{figure}


\begin{figure*}
[htb]
\begin{center}
\includegraphics[
width=0.65\textwidth
]%
{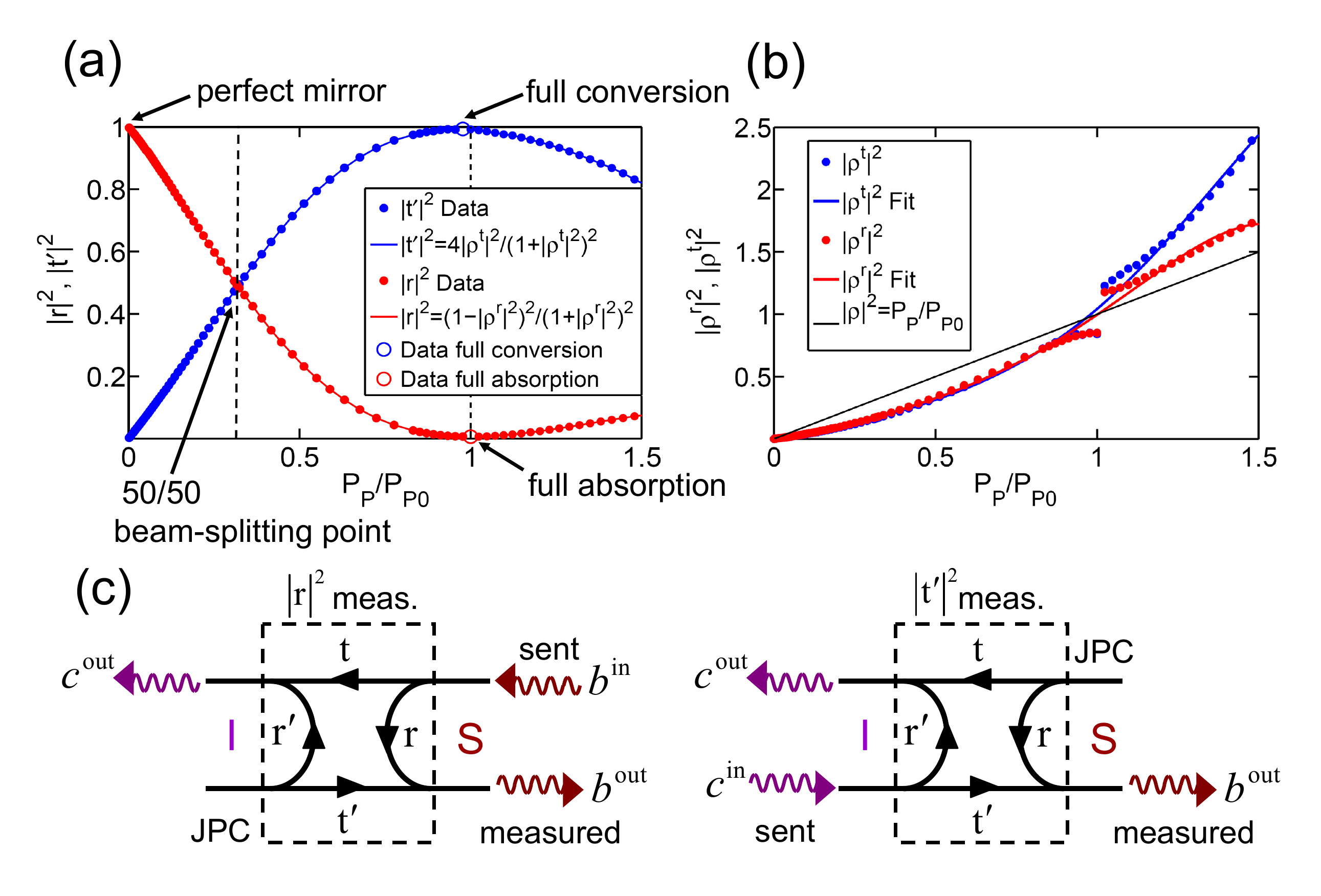}%
\caption{(color online). (a) Reflection and conversion measurements of the JPC in the conversion mode. The reflection parameter $\left\vert r\right\vert ^{2}$ and
the conversion parameter $\left\vert t^{\prime}\right\vert ^{2}$, drawn
as filled red and blue circles respectively, are measured at the signal port
as a function of the normalized applied pump power $P_{P}/P_{P0}$. In the
reflection and conversion measurement a signal and an idler tone are
applied at $f_{S}$ and $f_{I}$ respectively as shown in panel (c). The input photon flux
applied in each beam was set to be equal at the $50/50$ beam-splitting point as
explained in the text. In both measurements, the pump frequency is set to
$f_{P}=f_{I}-f_{S}$. The solid red and blue lines correspond to
theory expressions. The vertical dashed line on the left (right) indicates the
$50/50$ beam-splitting (full absorption) working point. In panel (b) we plot
using red (blue) circles the relation between the dimensionless parameter
$\left\vert \rho^{r}\right\vert ^{2}$ ($\left\vert \rho^{t}\right\vert ^{2}$)
employed in the fits of the reflection (conversion) data (shown in panel
(a)) and the normalized pump power applied in the experiment. The solid black
line of unity slope corresponds to the expected dependence of $\left\vert
\rho\right\vert ^{2}$ on $P_{P}/P_{P0}$ based on an ideal model of the JPC and
stiff pump condition. The solid red and blue curves are fifth order
polynomial fits. (c) Signal flow graphs of the JPC in the conversion
mode. In the reflection measurement (shown on the left) a coherent
beam is applied to the signal port, whereas in the conversion measurement
(shown on the right) an equivalent coherent beam is applied to the
idler ports. In both measurements, the output field is
measured at the signal port. }%
\label{RT}%
\end{center}
\end{figure*}

To introduce the requirements obeyed by the properties of unitary frequency conversion, we compare, in Fig. \ref{BS}, a 3-wave coherent converter and a 2-wave coherent beam-splitter. Panel (a) depicts the desired case of unitary full frequency conversion of an incident propagating beam, i.e. a violet (high frequency) to a red (low frequency) beam, obtained by pumping a dispersive $\chi_{2}$ medium with a yellow beam whose frequency is precisely the frequency difference. In panel (b) we depict the 2-wave analogue of such a device, that is the total reflector (mirror) whose input and output beams have the same frequency and magnitude but propagate in different directions. Panels (c) and (e) depict the same 3-wave coherent converter shown in panel (a), now operated at the 50/50 beam-splitting point. At this point, half of the power of input beams (red and violet) is transmitted while the other half undergoes frequency conversion, again using the third yellow beam. As can be seen in panels (c) and (e), either destructive or constructive 3-wave interference can appear at the output ports of the device depending on the relative phases of the two equal coherent incident red and violet beams. More specifically, panel (c) (panel (e)) describes a destructive interference scenario for the output red (violet) beam. Panels (c) and (e) also indicate the notions that, (1) the output beams of such a device depend on the phases of all incident coherent beams, (2) in order for a lossless destructive interference to take place between two frequencies, the power must be entirely converted to a third frequency. For further clarification of the role of the relative phases, we also show in panels (d) and (f) the 2-wave analogues of the destructive interference scenarios shown in panels (c) and (e).
 
Akin to the conceptual 3-wave converter depicted in Fig. \ref{BS}, the JPC is a non-degenerate device with spatial and temporal separation
between the idler and signal modes. Its input and output fields share however the same spatial port as shown in Fig. \ref{setup} (a) and must therefore be separated using a circulator. The idler and signal modes of the JPC are differential modes of the microstrip resonators of the device which intersect at a Josephson ring modulator (JRM) \cite{Jamp}. An optical micrograph of the JRM, which consists of four Josephson junctions, is shown in Fig. \ref{setup} (b). The third mode supported by the device is a non-resonant common-mode drive (pump), whose frequency $f_{P}$ is set to either the sum of the idler $f_{I}$ and signal $f_{S}$
frequencies or their difference. In the case where the pump frequency is equal
to the sum $f_{P}=f_{I}+f_{S}$, the device serves as a quantum-limited
amplifier which can be used to readout the state of a solid state qubit in
real time as has been demonstrated recently \cite{QuantumJumps,QubitJPC}. In
the present case where the pump frequency verifies $f_{P}%
=f_{I}-f_{S}$ ($f_{I}>f_{S}$), the device operates in the mode of frequency conversion with no photon gain. In this mode, as
opposed to amplification \cite{JPCnature}, the device is not required,
according to Caves theorem \cite{Caves} to add any noise. Note that this
working regime has been recently the subject of several works in the areas of telecommunication and
quantum information processing in optics \cite{VisibleZaske,Ates,QuantumRakher,PhotonicInt,IntTwoPhoton,CPC}. Also, in the recent work done at NIST \cite{QSJAumentado,QIJAumentado}, the authors  parametrically converted photons of different frequencies inside a microwave resonator coupled to a dc-SQUID. Phonon-photon parametric frequency conversion is also at play in the active cooling of a micro- or nano- mechanical resonator modes \cite{Painter,Teufel,kippenberg}. In our work, by contrast, it is photons
from two different spatial and temporal modes that are interconverted.

When operated in conversion mode and under the stiff pump condition, the JPC can be described as an effective two-port beam-splitter whose scattering parameters can be adjusted by varying the pump tone of the device. In this mode of operation, part of the incoming wave
at the idler (signal) port is transmitted to the signal (idler) port after being downconverted (upconverted), via emission (absorption) of pump photons, while the remaining part is reflected off the idler (signal) port. 

In Fig. \ref{RT} (a) we display measurements of the reflection parameter
$\left\vert r\right\vert ^{2}$ at the signal port (filled red circles) and the idler-to-signal conversion
parameter $\left\vert t^{\prime}\right\vert ^{2}$ (filled blue circles) as a function of the
normalized applied pump power $P_{P}/P_{P0}$, where $P_{P0}$ is the pump power
at which $\left\vert r\right\vert ^{2}$ is minimum. The frequency of
the pump tone in both measurements is $f_{P}=6.285$ $%
\operatorname{GHz}%
$. In the reflection measurement, a coherent tone at $f_{S}=8.402$ $%
\operatorname{GHz}%
$ is applied to the signal port with input power $P_{S0}=-123$ dBm, while in
the conversion measurement a coherent tone at $f_{I}=14.687$
$%
\operatorname{GHz}%
$ is applied to the idler port (see signal flow graphs of the JPC shown in
panel (c)). Both measurements are taken using a spectrum analyzer centered at
$f_{S}$ in zero frequency span mode. The output power measured as a function of
$P_{P}/P_{P0}$ is normalized relative to the reflected signal power obtained with no applied pump power. There, the JPC has unity reflection, which can be measured within $\pm0.5$ dB accuracy due to a finite impedance mismatch in the output line.   

As can be seen in Fig.  \ref{RT}, the splitting ratio between the converted and reflected portions of the beam is set by the device parameters and the pump amplitude. By varying the intensity of the pump, the splitting
ratio of the device can be changed continuously from zero conversion
and total reflection, to complete conversion and perfect
absorption (full converter). It is important to point out that the input
power applied to the idler port in the conversion measurement is set to
yield, at the $50/50$ beam-splitting working point, the same signal output power
as the one measured in reflection for $P_{S0}$. It is remarkable that
with this single calibration of input powers, which balances the output at the
$50/50$ beam-splitting point, the device satisfies quite well the equation of
conservation of total photon number $\left\vert r\right\vert ^{2}+\left\vert
t\right\vert ^{2}=1$ ($\left\vert t\right\vert =\left\vert t^{\prime
}\right\vert$) in the range $P_{P}/P_{P0}<1$. When the critical power is traversed $P_{P}/P_{P0}\gtrsim1$, we observe a progressive breaking of unitarity $\left\vert r\right\vert ^{2}+\left\vert t^{\prime}\right\vert ^{2}<1$. This can be
explained by increased nonlinear effects at elevated pump powers, resulting in
frequency conversion to higher modes of the system. However, we find that in the vicinity of $P_{P0}$, the conversion loss, defined as the deviation of $\left\vert r\right\vert ^{2}+\left\vert t^{\prime}\right\vert ^{2}$ from unity, to be less than $ \pm 10^{-2}$ within $ \pm 2.5\%$ accuracy. We also observe a slight shift in power between the peak of the
conversion and the minimum of the reflection data (indicated by empty
circles). The solid blue and red curves drawn using the equations shown in the legend of
Fig. \ref{RT} (a), correspond to theory expressions to the conversion and
reflection data respectively (see section III in the supplementary material). In panel (b) we plot, using filled red and blue
circles, the dimensionless pump power parameters $\left\vert \rho^{r}\right\vert ^{2}$ and
$\left\vert \rho^{t}\right\vert ^{2}$, which are used to fit the reflection and conversion data as a function of $P_{P}/P_{P0}$. The solid black line satisfying the relation $\left\vert \rho\right\vert ^{2}=$ $P_{P}/P_{P0}$, corresponds to the expected dependence of $\left\vert \rho\right\vert ^{2}$ on
$P_{P}/P_{P0}$ for the ideal JPC model with stiff pump approximation. The solid red and blue
curves are fifth order polynomial fits to the $\left\vert \rho^{r}\right\vert
^{2}$ and $\left\vert \rho^{t}\right\vert ^{2}$ parameters. The fits satisfy three
important properties: (1) they pass through the origin ($\left\vert \rho
^{r,t}\right\vert ^{2}=0$ for $P_{P}=0$), (2) they have a leading linear term in the
limit of small pump powers ($P_{P}\rightarrow0$), (3) the fits for $\left\vert
\rho^{r}\right\vert ^{2}$ and $\left\vert \rho^{t}\right\vert ^{2}$ coincide
for $P_{P}/P_{P0}<1$. 


\begin{figure}
[t!]
\begin{center}
\includegraphics[
width=0.49\textwidth
]%
{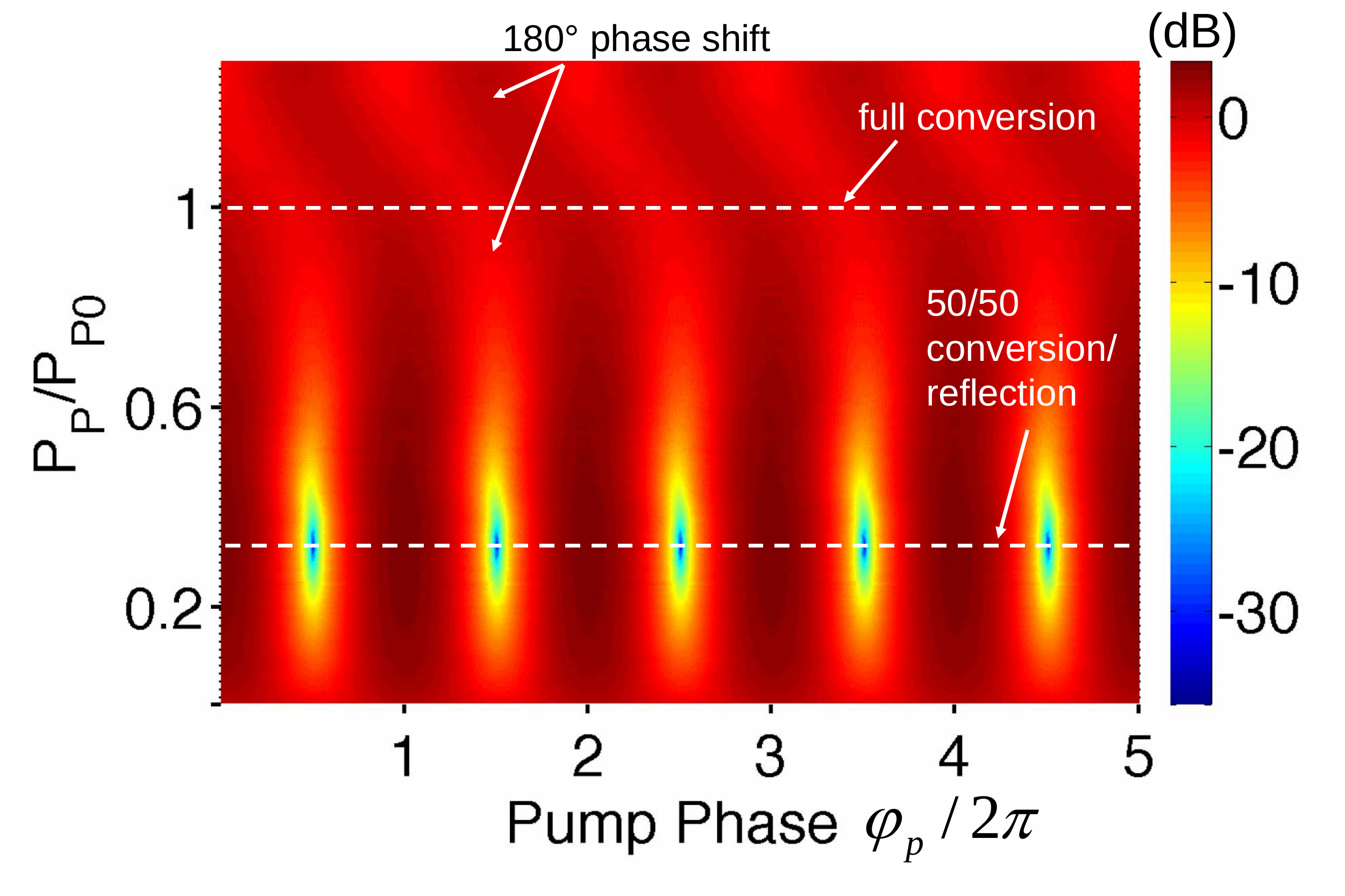}%
\caption{(color online). Interference fringes of the reflected signal and the converted idler
measured at the output of the signal port, at $f_{S}$, as a function of pump power and phase. Regions of low contrast
correspond to working points with unbalanced splitting ratios, namely, large
reflection $\left\vert r\right\vert ^{2}\rightarrow1$ or large conversion $\left\vert t^{\prime}\right\vert ^{2}\rightarrow1$. The highest
contrast, indicated by the bottom dashed white line, signals the 50/50
beam-splitting working point, where $\left\vert r\right\vert ^{2}=\left\vert
t^{\prime}\right\vert ^{2}=0.5$. The top dashed white line indicates the full conversion
working point where the interference contrast is lowest. For $P_{P}/P_{P0}>1$
($\left\vert \rho\right\vert ^{2}>1$) a phase shift is observed in the
interference pattern due to the change in the sign of the reflected signal
amplitude, in accordance with theory. The interference power is measured in dB
relative to the signal output power with no applied pump and idler tones.}%
\label{varypump}%
\end{center}
\end{figure}


\begin{figure}
[h!]
\begin{center}
\includegraphics[
width=0.49\textwidth
]%
{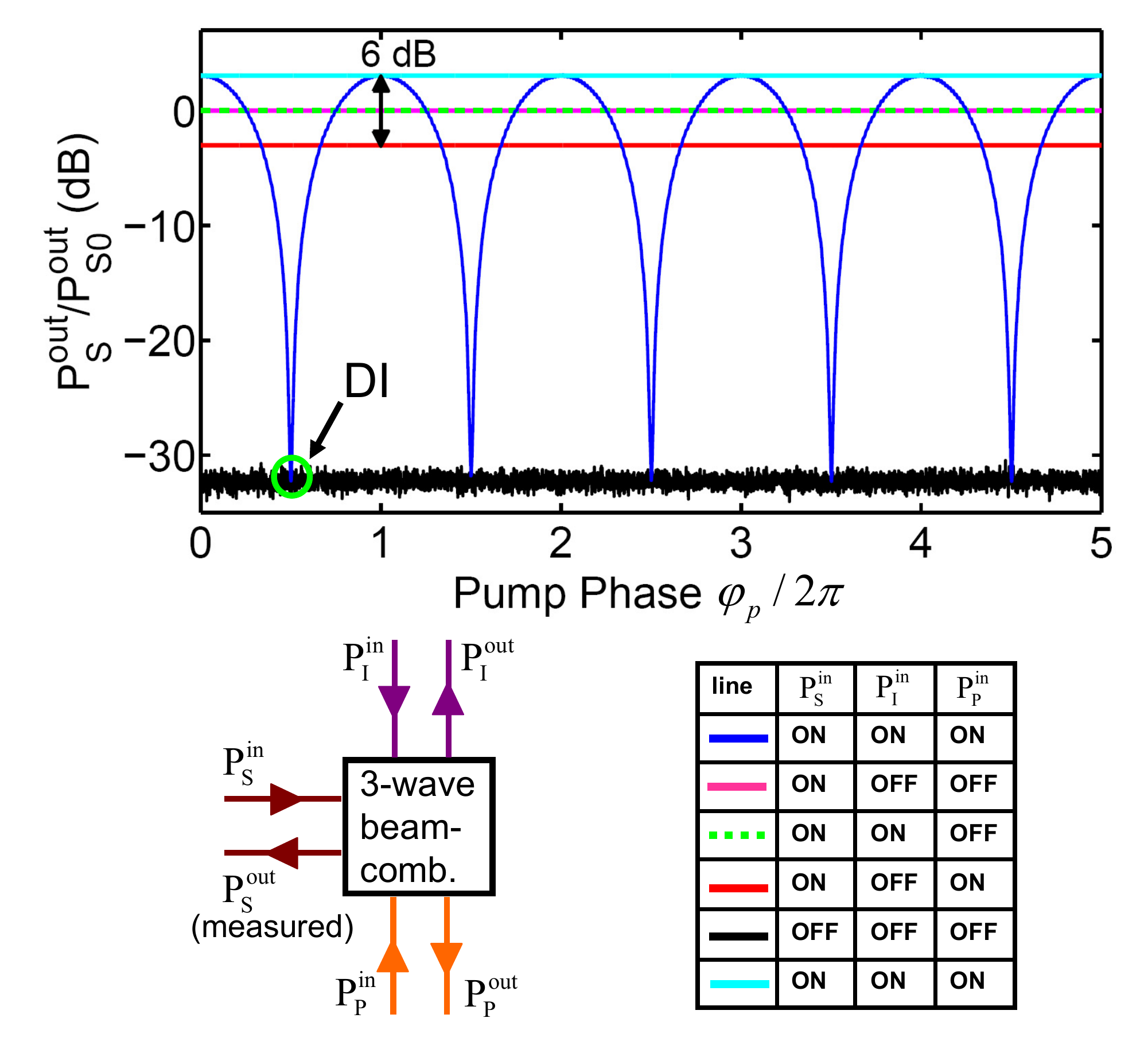}%
\caption{(color online). The blue curve is a cross-section of
the interference fringes shown in Fig. \ref{varypump} taken at the $50/50$
beam-splitting working point. The maximum and minimum points of the curve correspond to constructive and destructive
interference (DI) conditions respectively. The magenta line corresponds to the
reflected signal power $P_{S0}^{\mathrm{out}}$ with no applied pump and idler
tones (see the legend table and the device cartoon at the bottom). All curves
in the figure are plotted in dB relative to $P_{S0}^{\mathrm{out}}$. The green line corresponds to the relative reflected signal power with applied signal and idler
tones (pump off). The red line corresponds to the reflected signal power measured at the $50/50$ beam-splitting working
point (without idler). The black line corresponds to the relative noise floor
of the system set by the HEMT amplifier, measured with no applied tones and as
a function of time and not pump phase. The cyan line corresponds to a constructive
interference measurement taken for a certain pump phase.}%
\label{wi}%
\end{center}
\end{figure}

In addition to the device dependence on pump power shown in Fig. \ref{RT}, the
relative pump phase plays an important role as well. The downconverted
idler and the upconverted signal acquire a phase shift which depends on the
pump phase. We utilize this phase dependence as shown in Fig. \ref{varypump} in order to sensitively interfere signal and idler beams. In this measurement, we applied two coherent tones to the
signal and idler ports, having the same frequencies and powers as in the
measurement of Fig. \ref{RT}. We have also phase-locked the three independent coherent waves to the $10$ 
$\operatorname{MHz}$ reference oscillator of a rubidium atomic clock. The color mesh in Fig. \ref{varypump} depicts
the interference fringes of the reflected and converted signals generated
at the signal port as a function of $P_{P}/P_{P0}$ and the relative pump
phase. As can be seen in the figure, the interference contrast is low for $P_{P}\ll
P_{P0}$ which correspond to almost total reflection.
Similarly, the contrast is low for $P_{P}/P_{P0}$ close to unity which
correspond to almost total conversion. A maximum interference contrast
is obtained on the other hand for $P_{P}/P_{P0}$ close to $0.32$. We interpret this point as that where our device functions as a $50$/$50$ beam-splitter/combiner. This is further shown in Fig. \ref{wi}, where we display a cross-section of Fig. \ref{varypump} at the point of maximum interference contrast. The wave interference modulation curve, shown in blue, is plotted as a function of the relative phase of the pump. The other colored lines represent different
reference measurements taken under the same experimental conditions. The
magenta line corresponds to the reflected signal power $P_{S0}^{\mathrm{out}}$ obtained without pump or idler tones and forms the reference level for
the other measured powers. The dashed green line corresponds to the reflected signal
power obtained with no pump but with an equivalent idler tone injected through
the idler port. The fact that the dashed green line coincides with the magenta line
shows that without pump the idler and signal modes are isolated. The red line represents the
reflected signal power at the $50$/$50 $ beam-splitting working point and it
lies $-3$ dB below the signal reference line (magenta) as expected. The cyan
line corresponds to a maximum constructive interference of the
reflected signal and the converted idler, obtained for a certain pump
phase. As expected in this constructive interference condition, the relative power level is $6$ dB
(a factor of $4$) above the red line obtained without idler. The black line corresponds to the noise floor measured without any
signal. This noise is dominated by the high electron mobility transistor
(HEMT) amplifier noise connected at the $4%
\operatorname{K}%
$ stage of our setup. Note that the minimum points of
the wave interference curve (blue) correspond to destructive
interference (DI) condition and coincide with the HEMT noise floor. By calculating the ratio of the average output power received in
a destructive interference experiment to the average power received in a
constructive interference experiment, as discussed in section IV in the supplementary
material, we get a lower bound (defined for the absolute value) of $-37.9$ dB
on the amount of destructive interference achieved by the device at the
$50/50$ beam-splitting point. This is equivalent to a coherent cancellation
of $99.98\%$ of input signal and idler beams.

Furthermore, in a separate interference measurement in which we varied the input power of the signal and idler beams (see supplementary material), we find that the maximum input power which can be handled by the
device is about $-100$ dBm. Similar to Josephson amplifiers based on the JPC,
this figure of merit can be improved by about one order of magnitude,
by increasing the critical current of the junctions as discussed in Refs.
\cite{Jamp,JRCAreview}. We also note that the bandwidth of the present device
($\sim70
\operatorname{MHz}
$) is limited by the bandwidths of the resonators. A wider bandwidth, on the order of $500%
\operatorname{MHz}
$, could potentially be achieved by substituting the present JRM design with an inductively
shunted ring \cite{Roch}.

In addition to quantum information transduction, the device has several potential useful applications, such as cooling of a readout
cavity of a qubit by swapping the ``hot'' cavity photons, for instance at $8$ $%
\operatorname{GHz}%
$, by ``cold''  reservoir photons at $15$ $%
\operatorname{GHz}%
$, and realization of a Mach-Zehnder interferometer (MZI) scheme for microwaves with real-time
feedback. In the latter scheme, the device would function as an interferometric $50/50$
beam-combiner for incoming idler and signal waves. Assuming, for example, the information is carried by the phase of the signal in the MZI setup, while the idler wave serves as a reference, the information can be decoded from the incoming signal by measuring the generated interference at the signal port. Moreover, the setpoint of the device, which
yields maximum phase sensitivity, can be maintained in-situ by compensating the shift using the pump phase. 

Discussions with R. J. Schoelkopf and P. T. Rakich are gratefully acknowledged. The assistance
of Michael Power in the fabrication process is highly appreciated. This
research was supported by the NSF under grants DMR-1006060 and DMR-0653377; ECCS-1068642, the
NSA through ARO Grant No. W911NF-09-1-0514, IARPA under ARO Contract No.
W911NF-09-1-0369, the Keck foundation, and Agence Nationale pour la Recherche
under grant ANR07-CEXC-003. M.H.D. acknowledges partial support from College
de France.

\end{document}


\title{Supplementary Material for ``Full coherent frequency conversion between two microwave propagating modes''}
\author{Baleegh Abdo}
\email{baleegh.abdo@yale.edu}
\author{Katrina Sliwa}
\author{Flavius Schackert}
\author{Nicolas Bergeal}
\altaffiliation{Current address: LPEM-UMR8213/CNRS-ESPCI ParisTech-UPMC, 10 rue Vauquelin-75005 Paris, France.}
\author{Michael Hatridge}
\author{Luigi Frunzio}
\author{A. Douglas Stone}
\author{Michel Devoret}
\affiliation{Department of Applied Physics, Yale University, New Haven, CT 06520, USA.}
\date{\today }

\maketitle

\section{Ideal mixer equations}

The Hamiltonian of an ideal 3-wave mixing device with internal modes
\textit{a}, \textit{b} and \textit{c} can be written in the framework of the
Rotating Wave Approximation (RWA) of the system as,%

\begin{equation}
\frac{H_{0}^{\mathrm{RWA}}}{\hbar}=\omega_{a}a^{\dag}a+\omega_{b}b^{\dag
}b+\omega_{c}c^{\dag}c+g_{3}\left(  a^{\dag}b^{\dag}c+abc^{\dag}\right)
,\label{H0}%
\end{equation}
where we kept only terms commuting with the total photon number. The operators
$a^{\dag}$, $b^{\dag}$, $c^{\dag}$ ($a$, $b$, $c$) are the creation
(annihilation) operators of modes \textit{a}, \textit{b} and \textit{c}
respectively, which satisfy the bosonic commutation relations $\left[
a,a^{\dag}\right]  =\left[  b,b^{\dag}\right]  =\left[  c,c^{\dag}\right]
=1$. The angular frequencies $\omega_{a}$, $\omega_{b}$, $\omega_{c}$ are the
resonance frequencies of modes \textit{a}, \textit{b} and \textit{c}, where
$\omega_{a}<$ $\omega_{b}<$ $\omega_{c}$ and $\omega_{c}$ satisfies
$\omega_{c}=\omega_{a}+\omega_{b}$. The angular frequency $g_{3}$ denotes the
coupling constant between the modes.

Treating the coupling of each oscillator with a transmission line carrying
waves in and out of the oscillator in the RWA, one arrives at three coupled
quantum Langevin equations
\begin{align}
\frac{\mathrm{d}}{\mathrm{d}t}a  &  =-i\omega_{a}a-ig_{3}b^{\dag}%
c-\frac{\gamma_{a}}{2}a+\sqrt{\gamma_{a}}\tilde{a}^{\mathrm{in}}\left(
t\right)  ,\nonumber\\
\frac{\mathrm{d}}{\mathrm{d}t}b  &  =-i\omega_{b}b-ig_{3}a^{\dag}%
c-\frac{\gamma_{b}}{2}b+\sqrt{\gamma_{b}}\tilde{b}^{\mathrm{in}}\left(
t\right)  ,\nonumber\\
\frac{\mathrm{d}}{\mathrm{d}t}c  &  =-i\omega_{c}c-ig_{3}ab-\frac{\gamma_{c}%
}{2}c+\sqrt{\gamma_{c}}\tilde{c}^{\mathrm{in}}\left(  t\right)
,\label{threeAmpEqs}%
\end{align}
where the second term in the right hand side corresponds to the non-linear
term producing photon conversion, $\gamma_{a,b,c}$ is the rate at which
photons introduced in one resonator leave, and $\tilde{a}^{\mathrm{in}}\left(
t\right)  $, $\tilde{b}^{\mathrm{in}}\left(  t\right)  $, $\tilde
{c}^{\mathrm{in}}\left(  t\right)  $ are input fields which correspond to the
negative frequency component of the drive terms in the classical equations.
Taking for example $\tilde{a}^{\mathrm{in}}\left(  t\right)  $, the input
fields are given by
\begin{equation}
\tilde{a}^{\mathrm{in}}(t)=\frac{1}{\sqrt{2\pi}}\int_{0}^{+\infty
}a^{\mathrm{in}}\left[  \omega\right]  e^{-i\omega t}\mathrm{d}\omega,
\end{equation}
where $a^{\mathrm{in}}[\omega]$ are the usual field operators obeying the
commutation relations
\begin{equation}
\left[  a^{\mathrm{in}}\left[  \omega\right]  ,a^{\mathrm{in}}\left[
\omega^{\prime}\right]  \right]  =\mathrm{sgn}\left(  \frac{\omega
-\omega^{\prime}}{2}\right)  \delta\left(  \omega+\omega^{\prime}\right)
\end{equation}
in which $\omega$ denotes a frequency that can be positive or negative.

Using the standard input-output relations for the three oscillators given by%
\begin{align}
\sqrt{\gamma_{a}}a\left(  t\right)   &  =\tilde{a}^{\mathrm{in}}\left(
t\right)  +\tilde{a}^{\mathrm{out}}\left(  t\right)  ,\label{InOuta}\\
\sqrt{\gamma_{b}}b\left(  t\right)   &  =\tilde{b}^{\mathrm{in}}\left(
t\right)  +\tilde{b}^{\mathrm{out}}\left(  t\right)  ,\label{InOutb}\\
\sqrt{\gamma_{c}}c\left(  t\right)   &  =\tilde{c}^{\mathrm{in}}\left(
t\right)  +\tilde{c}^{\mathrm{out}}\left(  t\right)  ,\label{InOutc}%
\end{align}
we can express equations (\ref{threeAmpEqs}) as%

\begin{align}
O_{a}^{+}\tilde{a}^{\mathrm{out}}\left(  t\right)   &  =-O_{a}^{-}\tilde
{a}^{\mathrm{in}}\left(  t\right)  -i\sqrt{\gamma_{a}}g_{3}b^{\dag
}c,\label{a_t_eq}\\
O_{b}^{+}\tilde{b}^{\mathrm{out}}\left(  t\right)   &  =-O_{b}^{-}\tilde
{b}^{\mathrm{in}}\left(  t\right)  -i\sqrt{\gamma_{b}}g_{3}a^{\dag
}c,\label{b_t_eq}\\
O_{c}^{+}\tilde{c}^{\mathrm{out}}\left(  t\right)   &  =-O_{c}^{-}\tilde
{c}^{\mathrm{in}}\left(  t\right)  -i\sqrt{\gamma_{c}}g_{3}ab,\label{c_t_eq}%
\end{align}
where%

\begin{align}
O_{a,b,c}^{\pm}  &  =\frac{\mathrm{d}}{\mathrm{d}t}+i(\omega_{a,b,c}\mp
i\Gamma_{a,b,c}),\label{BigO}\\
\Gamma_{a,b,c}  &  =\frac{\gamma_{a,b,c}}{2}.\label{Gamma}%
\end{align}
Equations (\ref{a_t_eq}), (\ref{b_t_eq}), (\ref{c_t_eq}) in combination with
relations (\ref{InOuta}), (\ref{InOutb}), (\ref{InOutc}) form a set of $6$
coupled equations which represent an ideal 3-wave mixing device.
Unfortunately, these equations lack an analytical solution in the general case
where all modes and drives are treated on the same footing. However, an
analytical solution can be found under certain conditions as discussed in the
following subsections, namely, if all drives are very small or if one of the
drives is stronger than the others.

\section{Small signal approximation}

Fourier transforming equations (\ref{a_t_eq}), (\ref{b_t_eq}), (\ref{c_t_eq}) gives%

\begin{align}
\chi_{a}^{-1}\left[  \omega_{1}\right]  a^{\mathrm{out}}\left[  \omega
_{1}\right]   &  =\chi_{a}^{-1\ast}\left[  \omega_{1}\right]  a^{\mathrm{in}%
}\left[  \omega_{1}\right]  -i\frac{2g_{3}}{\sqrt{\gamma_{a}}}\left(  b^{\dag
}\left[  \omega_{2}\right]  \ast c\left[  \omega_{3}\right]  \right)
,\label{a_w_eq}\\
\chi_{b}^{-1}\left[  \omega_{2}\right]  b^{\mathrm{out}}\left[  \omega
_{2}\right]   &  =\chi_{b}^{-1\ast}\left[  \omega_{2}\right]  b^{\mathrm{in}%
}\left[  \omega_{2}\right]  -i\frac{2g_{3}}{\sqrt{\gamma_{b}}}\left(  a^{\dag
}\left[  \omega_{1}\right]  \ast c\left[  \omega_{3}\right]  \right)
,\label{b_w_eq}\\
\chi_{c}^{-1}\left[  \omega_{3}\right]  c^{\mathrm{out}}\left[  \omega
_{3}\right]   &  =\chi_{c}^{-1\ast}\left[  \omega_{3}\right]  c^{\mathrm{in}%
}\left[  \omega_{3}\right]  -i\frac{2g_{3}}{\sqrt{\gamma_{c}}}\left(  a\left[
\omega_{1}\right]  \ast b\left[  \omega_{2}\right]  \right)  ,\label{c_w_eq}%
\end{align}
where the asterisk ``$\ast$" stands for convolution operation, $\omega_{1}$,
$\omega_{2}$, $\omega_{3}$ are the excitation angular frequencies of modes
\textit{a}, \textit{b} and \textit{c} respectively, $\chi_{a,b,c}$ is a
dimensionless response function whose inverse is given by%

\begin{equation}
\chi_{a,b,c}^{-1}\left[  \omega_{1,2,3}\right]  =1-i\frac{\omega
_{1,2,3}-\omega_{a,b,c}}{\Gamma_{a,b,c}},\label{ita_a_b_c}%
\end{equation}
and $a\left[  \omega_{1}\right]  $, $b\left[  \omega_{2}\right]  $, $c\left[
\omega_{3}\right]  $ satisfy the input-output relations in the frequency
domain which read%

\begin{align}
\sqrt{\gamma_{a}}a\left[  \omega_{1}\right]   &  =\tilde{a}^{\mathrm{in}%
}\left[  \omega_{1}\right]  +\tilde{a}^{\mathrm{out}}\left[  \omega
_{1}\right]  ,\label{a_w}\\
\sqrt{\gamma_{b}}b\left[  \omega_{2}\right]   &  =\tilde{b}^{\mathrm{in}%
}\left[  \omega_{2}\right]  +\tilde{b}^{\mathrm{out}}\left[  \omega
_{2}\right]  ,\label{b_w}\\
\sqrt{\gamma_{c}}c\left[  \omega_{3}\right]   &  =\tilde{c}^{\mathrm{in}%
}\left[  \omega_{3}\right]  +\tilde{c}^{\mathrm{out}}\left[  \omega
_{3}\right]  .\label{c_w}%
\end{align}
In case the drives are small $\left\vert \left\langle \tilde{a}^{\mathrm{in}%
}\right\rangle \right\vert ^{2}\ll1$, $\left\vert \left\langle \tilde
{b}^{\mathrm{in}}\right\rangle \right\vert ^{2}\ll1$, $\left\vert \left\langle
\tilde{c}^{\mathrm{in}}\right\rangle \right\vert ^{2}\ll1$, we can obtain a
first order solution for the internal modes of the system
\begin{align}
a\left[  \omega_{1}\right]   &  =2\chi_{a}\left[  \omega_{1}\right]
\frac{\tilde{a}^{\mathrm{in}}\left[  \omega_{1}\right]  }{\sqrt{\gamma_{a}}%
},\label{a_w_1}\\
b\left[  \omega_{2}\right]   &  =2\chi_{b}\left[  \omega_{2}\right]
\frac{\tilde{b}^{\mathrm{in}}\left[  \omega_{2}\right]  }{\sqrt{\gamma_{b}}%
},\label{b_w_2}\\
c\left[  \omega_{3}\right]   &  =2\chi_{c}\left[  \omega_{3}\right]
\frac{\tilde{c}^{\mathrm{in}}\left[  \omega_{3}\right]  }{\sqrt{\gamma_{c}}%
},\label{c_w_3}%
\end{align}
by neglecting the nonlinear terms on the right-hand side of equations
(\ref{a_w_eq}), (\ref{b_w_eq}), (\ref{c_w_eq}) and substituting the resultant
output fields in equations (\ref{a_w}), (\ref{b_w}), (\ref{c_w}).

Furthermore, by using equations (\ref{a_w_1}), (\ref{b_w_2}), (\ref{c_w_3}) as
perturbation in equations (\ref{a_w_eq}), (\ref{b_w_eq}), (\ref{c_w_eq}) we get%

\begin{align}
\chi_{a}^{-1}\left[  \omega_{1}\right]  \tilde{a}^{\mathrm{out}}\left[
\omega_{1}\right]   &  =\chi_{a}^{-1\ast}\left[  \omega_{1}\right]  \tilde
{a}^{\mathrm{in}}\left[  \omega_{1}\right]  -i\frac{8g_{3}}{\sqrt{\gamma
_{a}\gamma_{b}\gamma_{c}}}\left(  \frac{b^{\mathrm{in}}\left[  -\omega
_{2}\right]  }{\chi_{b}^{-1\ast}\left[  \omega_{2}\right]  }\ast
\frac{c^{\mathrm{in}}\left[  \omega_{3}\right]  }{\chi_{c}^{-1}\left[
\omega_{3}\right]  }\right)  ,\\
\chi_{b}^{-1}\left[  \omega_{2}\right]  \tilde{b}^{\mathrm{out}}\left[
\omega_{2}\right]   &  =\chi_{b}^{-1\ast}\left[  \omega_{2}\right]  \tilde
{b}^{\mathrm{in}}\left[  \omega_{2}\right]  -i\frac{8g_{3}}{\sqrt{\gamma
_{a}\gamma_{b}\gamma_{c}}}\left(  \frac{a^{\mathrm{in}}\left[  -\omega
_{1}\right]  }{\chi_{a}^{-1\ast}\left[  \omega_{1}\right]  }\ast
\frac{c^{\mathrm{in}}\left[  \omega_{3}\right]  }{\chi_{c}^{-1}\left[
\omega_{3}\right]  }\right)  ,\\
\chi_{c}^{-1}\left[  \omega_{3}\right]  \tilde{c}^{\mathrm{out}}\left[
\omega_{3}\right]   &  =\chi_{c}^{-1\ast}\left[  \omega_{3}\right]  \tilde
{c}^{\mathrm{in}}\left[  \omega_{3}\right]  -i\frac{8g_{3}}{\sqrt{\gamma
_{a}\gamma_{b}\gamma_{c}}}\left(  \frac{a^{\mathrm{in}}\left[  \omega
_{1}\right]  }{\chi_{a}^{-1}\left[  \omega_{1}\right]  }\ast\frac
{b^{\mathrm{in}}\left[  \omega_{2}\right]  }{\chi_{b}^{-1}\left[  \omega
_{2}\right]  }\right)  .
\end{align}
At resonance tuning $\omega_{1}=$ $\omega_{a}$, $\omega_{2}=\omega_{b}$,
$\omega_{3}=\omega_{c}$ where $\chi_{a,b,c}^{-1}=1$ this solution reduces into%

\begin{align}
\tilde{a}^{\mathrm{out}}\left[  \omega_{a}\right]   &  =\tilde{a}%
^{\mathrm{in}}\left[  \omega_{a}\right]  -i\frac{8g_{3}}{\sqrt{\gamma
_{a}\gamma_{b}\gamma_{c}}}\left(  b^{\mathrm{in}}\left[  -\omega_{b}\right]
\ast c^{\mathrm{in}}\left[  \omega_{c}\right]  \right)  ,\label{a_res_ss}\\
\tilde{b}^{\mathrm{out}}\left[  \omega_{b}\right]   &  =\tilde{b}%
^{\mathrm{in}}\left[  \omega_{b}\right]  -i\frac{8g_{3}}{\sqrt{\gamma
_{a}\gamma_{b}\gamma_{c}}}\left(  a^{\mathrm{in}}\left[  -\omega_{a}\right]
\ast c^{\mathrm{in}}\left[  \omega_{c}\right]  \right)  ,\label{b_res_ss}\\
\tilde{c}^{\mathrm{out}}\left[  \omega_{c}\right]   &  =\tilde{c}%
^{\mathrm{in}}\left[  \omega_{c}\right]  -i\frac{8g_{3}}{\sqrt{\gamma
_{a}\gamma_{b}\gamma_{c}}}\left(  a^{\mathrm{in}}\left[  \omega_{a}\right]
\ast b^{\mathrm{in}}\left[  \omega_{b}\right]  \right)  .\label{c_res_ss}%
\end{align}
Note that the convolution in the second term on the right hand side of
equations (\ref{a_res_ss}), (\ref{b_res_ss}), (\ref{c_res_ss}) represents a
product of the input fields in the time domain.

\section{JPC scattering matrix}

In this subsection, we derive the scattering matrix of the JPC\ in the noiseless conversion regime. Assuming that modes \textit{a} or \textit{b
}are strongly driven, which is the regime of operation discussed in this work,
then equations (\ref{a_t_eq}), (\ref{b_t_eq}), (\ref{c_t_eq}) can be solved
analytically and yield a scattering matrix of an ideal mixer. To obtain the
solution for this case, we assume, without loss of generality, that $\gamma
_{a}\gg\gamma_{b},\gamma_{c}$ and that we pump oscillator \textit{a }with a
strong classical drive. In this case, modes \textit{b }and\textit{\ c
}represent the signal and idler modes and \textit{a} the pump mode. Under
these conditions, we can replace operator $a\left(  t\right)  $ by the average
value in the coherent state produced by the pump $\left\langle a\left(
t\right)  \right\rangle =\left\vert \alpha\right\vert e^{-i\left(  \omega
_{p}t+\varphi_{p}\right)  }$, where $\left\vert \alpha\right\vert $ is the
amplitude of the coherent drive, $\varphi_{p}$ is the phase of the drive and
$\omega_{p}=\omega_{i}-\omega_{s}$, where $\omega_{s}$, $\omega_{i}$ are the
angular frequencies of the signal and idler excitations. Substituting this
relation in equations (\ref{a_t_eq}), (\ref{b_t_eq}), (\ref{c_t_eq}) and using
equations (\ref{InOuta}), (\ref{InOutb}), (\ref{InOutc}), one gets two
equations in the time domain for $\tilde{b}^{\mathrm{out}}\left(  t\right)  $
and $\tilde{c}^{\mathrm{out}}\left(  t\right)  $ as a function of $\tilde
{b}^{\mathrm{in}}\left(  t\right)  $ and $\tilde{c}^{\mathrm{in}}\left(
t\right)  $. By Fourier transforming these two equations into the frequency
domain, we arrive at the scattering matrix of the JPC in the conversion mode given by%

\begin{equation}
\left[
\begin{array}
[c]{c}%
b^{\mathrm{out}}\left[  \omega_{s}\right] \\
c^{\mathrm{out}}\left[  \omega_{i}\right]
\end{array}
\right]  =\left[
\begin{array}
[c]{cc}%
r_{bb} & t_{bc}\\
t_{cb} & r_{cc}%
\end{array}
\right]  \left[
\begin{array}
[c]{c}%
b^{\mathrm{in}}\left[  \omega_{s}\right] \\
c^{\mathrm{in}}\left[  \omega_{i}\right]
\end{array}
\right]  ,
\end{equation}
where $b^{\mathrm{in}}$, $c^{\mathrm{in}}$ and $b^{\mathrm{out}}$,
$c^{\mathrm{out}}$ are the incoming and outgoing wave amplitudes of the signal
and idler respectively, given in units of square root of photon number per
unit frequency, $r_{bb}$, $r_{cc}$ are the amplitude reflection parameters for
the signal and idler and $t_{bc}$, $t_{cb}$ are the conversion
parameters from idler to signal and vice versa, which are given by%

\begin{align}
r_{bb}  &  =\frac{\chi_{b}^{-1\ast}\chi_{c}^{-1}-\left\vert \rho\right\vert
^{2}}{\chi_{b}^{-1}\chi_{c}^{-1}+\left\vert \rho\right\vert ^{2}}%
,\label{rbb}\\
r_{cc}  &  =\frac{\chi_{b}^{-1}\chi_{c}^{-1\ast}-\left\vert \rho\right\vert
^{2}}{\chi_{b}^{-1}\chi_{c}^{-1}+\left\vert \rho\right\vert ^{2}}%
,\label{rcc}\\
t_{bc}  &  =\frac{2i\rho}{\chi_{b}^{-1}\chi_{c}^{-1}+\left\vert \rho
\right\vert ^{2}},\label{tbc}\\
t_{cb}  &  =\frac{2i\rho^{\ast}}{\chi_{b}^{-1}\chi_{c}^{-1}+\left\vert
\rho\right\vert ^{2}},\label{tcb}%
\end{align}
where the dimensionless response functions $\chi_{b,c}$ and the parameter
$\rho$ are given by%

\begin{align}
\chi_{b,c}^{-1} &  =1-i\frac{\omega_{s,i}-\omega_{b,c}}{\Gamma_{b,c}%
},\label{ita}\\
\rho &  =\sqrt{\frac{P_{P}}{P_{P0}}}e^{-i\varphi_{p}},\label{rho}%
\end{align}
where $P_{P}=$ $\hbar\omega_{p}\gamma_{p}\left\vert \alpha\right\vert ^{2}$ is
the pump power and $P_{P0}=\hbar\omega_{p}\gamma_{p}\Gamma_{b}\Gamma_{c}%
/g_{3}^{2}$ where $\gamma_{p}=\gamma_{a}$.%

Note that the scattering matrix is unitary (the total number of photons is
conserved) and satisfies the relations $\left\vert r_{bb}\right\vert
^{2}+\left\vert t_{cb}\right\vert ^{2}=1$, $\left\vert r_{cc}\right\vert
^{2}+\left\vert t_{bc}\right\vert ^{2}=1$.

For zero frequency detuning, i.e. $\chi_{b}^{-1}=\chi_{c}^{-1}=1$, the
scattering matrix parameters reduce to $r=r_{bb}=r_{cc}$, $t=t_{bc}%
=-t_{cb}^{\ast}$ where%

\begin{align}
r  &  =\frac{1-\left\vert \rho\right\vert ^{2}}{1+\left\vert \rho\right\vert
^{2}},\label{r}\\
t  &  =\frac{2i\rho}{1+\left\vert \rho\right\vert ^{2}}.\label{t}%
\end{align}
Equivalently, the unitary scattering matrix of the device can be written in
the form%

\begin{equation}
\left[
\begin{array}
[c]{cc}%
\cos\theta & e^{-i\varphi_{p}}\sin\theta\\
-e^{i\varphi_{p}}\sin\theta & \cos\theta
\end{array}
\right]  .
\end{equation}
It is worthwhile mentioning that if instead of oscillator \textit{a} and
\textit{b} discussed above, oscillator \textit{c }satisfies $\gamma_{c}\gg$
$\gamma_{a},\gamma_{b}$ and it is driven by a bright classical tone
$\left\vert \left\langle c^{\mathrm{in}}\right\rangle \right\vert ^{2}\gg1$,
which produces a coherent state, then equations (\ref{a_t_eq}), (\ref{b_t_eq}%
), (\ref{c_t_eq}) and (\ref{InOuta}), (\ref{InOutb}), (\ref{InOutc}) can be
solved in a similar manner as was done for modes \textit{a} and \textit{b} and
yield the scattering matrix of a quantum-limited amplifier (see Ref.
\cite{JRCAreview}).

\begin{figure}
[h]
\begin{center}
\includegraphics[
width=0.65\textwidth
]%
{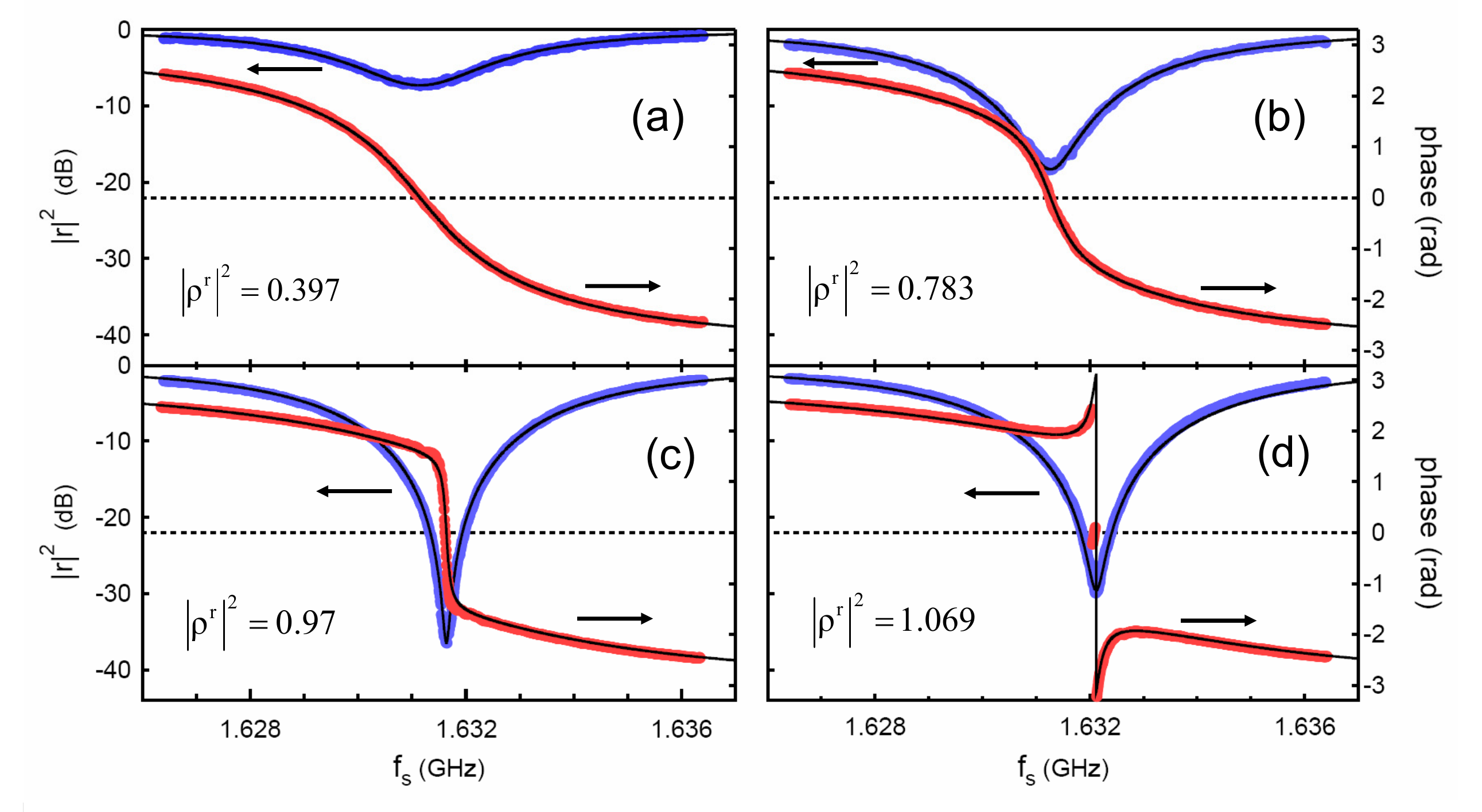}%
\caption{(colored online). Reflection parameter versus signal frequency. Reflection
parameter magnitude (blue curves)\ and phase (red curves) of a JPC device
measured in conversion mode as a function of the signal frequency.
The measurements exhibited in subplots (a), (b), (c) and (d) are taken using a
vector network analyzer and correspond to different applied pump powers. The
black curves are theory fits to the data using Eq. (\ref{rbb}). The values for
$\left\vert \rho^{r}\right\vert ^{2}$ listed in each panel are extracted from
the theory fits. The pump frequency applied in this measurement is
$f_{P}=5.585\operatorname{GHz}$. The resonance frequency of the idler mode is
$f_{b}=7.216\operatorname{GHz}$ and the quality factors of resonator
\textit{a} (S) and \textit{b }(I)\textit{\ }are $Q_{a}=450$ and $Q_{b}=120$
respectively.}%
\label{rf}%
\end{center}
\end{figure}

In Fig. \ref{rf} we show a reflection parameter measurement of a JPC operated
in conversion mode. The measured device was previously presented in
Ref. \cite{JPCnature} and operated as a quantum-limited amplifier. Subplots
(a), (b), (c) and (d) display reflection parameter data taken for different
applied pump powers at $f_{P}=5.585%
\operatorname{GHz}%
$. The blue and red curves correspond to the reflection parameter magnitude
and phase measured using a vector network analyzer as a function of the device
signal frequency. The black curves are theory fits to the data plotted using
equation (\ref{rbb}). As expected, the reflection magnitude varies with the
pump power. It decreases as the pump power is increased as shown in subplots
(a) and (b). It reaches a minimum for a certain pump power $P_{P0}$
corresponding to $\rho^{r}\simeq1^{-}$ as shown in subplot (c), whereas for
larger pump powers corresponding to $\rho^{r}>1$ it increases again as
displayed in subplot (d). Furthermore, as can be seen in subplot (d), the
reflection phase undergoes a sign change for $\rho^{r}>1$ as predicted by
Eq. (\ref{rbb}).

\subsection{Interference using the JPC}

In an interference experiment, two equal tones are injected to the signal and
idler ports of the JPC, that is $\left\vert b^{\mathrm{in}}\left[  \omega
_{s}\right]  \right\vert ^{2}=\left\vert c^{\mathrm{in}}\left[  \omega
_{i}\right]  \right\vert ^{2}$. The output signal photon number per unit
frequency reads%

\begin{align}
\left\vert b^{\mathrm{out}}\left[  \omega_{s}\right]  \right\vert ^{2}  &
=\left\vert rb^{\mathrm{in}}\left[  \omega_{s}\right]  +tc^{\mathrm{in}%
}\left[  \omega_{i}\right]  \right\vert ^{2}\\
&  =\left\vert \left\vert r\right\vert +\left\vert t\right\vert e^{i\left(
\varphi_{i}-\varphi_{s}-\varphi_{p}\right)  }\right\vert ^{2}\left\vert
b^{\mathrm{in}}\left[  \omega_{s}\right]  \right\vert ^{2},
\end{align}
where $\varphi_{s}$ and $\varphi_{i}$ are the phases of the input signal and
idler coherent signals.

At the $50/50$ beam-splitting point, $\left\vert r\right\vert =\left\vert
t\right\vert =1/\sqrt{2}$ which yields an interference pattern given by%

\begin{equation}
\frac{\left\vert b^{\mathrm{out}}\left[  \omega_{s}\right]  \right\vert ^{2}%
}{\left\vert b^{\mathrm{in}}\left[  \omega_{s}\right]  \right\vert ^{2}%
}=2\left\vert \cos\left(  \left(  \varphi_{i}-\varphi_{s}-\varphi_{p}\right)
/2\right)  \right\vert ^{2},\label{interference_relation}%
\end{equation}
which depends on the phases of the three coherent beams. When comparing the
peaks of these interference fringes to the reflection and conversion
parameters at the $50/50$ beam-splitting point ($\left\vert r\right\vert
^{2}=\left\vert t\right\vert ^{2}=1/2$) we obtain the $6$ dB difference
observed in the experiment (see Fig. 5 in the main text).

\subsection{Manley-Rowe relations}

In a reactive lossless nonlinear device with multiple frequencies such as the
JPC, the net flux of photons (number of photons per unit time) of pump $\Delta
N_{P}$, signal $\Delta N_{S}$ and idler $\Delta N_{I}$ participating in the
mixing interaction obey the following Manley-Rowe relations%

\begin{equation}
\left\vert \Delta N_{P}\right\vert =\left\vert \Delta N_{S}\right\vert
=\left\vert \Delta N_{I}\right\vert ,\label{MRrelation}%
\end{equation}
where
\begin{align}
\left\vert \Delta N_{P}\right\vert  &  =\left\vert N_{P}^{\mathrm{in}}%
-N_{P}^{\mathrm{out}}\right\vert ,\\
\left\vert \Delta N_{S}\right\vert  &  =\left\vert N_{S}^{\mathrm{in}}%
-N_{S}^{\mathrm{out}}\right\vert ,\\
\left\vert \Delta N_{I}\right\vert  &  =\left\vert N_{I}^{\mathrm{in}}%
-N_{I}^{\mathrm{out}}\right\vert ,
\end{align}
and $N_{P}^{\mathrm{in,out}}=\left\vert \tilde{a}^{\mathrm{in,out}}\left(
t\right)  \right\vert ^{2}$, $N_{S}^{\mathrm{in,out}}=\left\vert \tilde
{b}^{\mathrm{in,out}}\left(  t\right)  \right\vert ^{2}$, $N_{I}%
^{\mathrm{in,out}}=\left\vert \tilde{c}^{\mathrm{in,out}}\left(  t\right)
\right\vert ^{2}$. Equation (\ref{MRrelation}) can be also written in terms of
the net power contained in the different modes%

\begin{equation}
\left\vert \frac{\Delta P_{P}}{\hbar\omega_{p}}\right\vert =\left\vert
\frac{\Delta P_{S}}{\hbar\omega_{s}}\right\vert =\left\vert \frac{\Delta
P_{I}}{\hbar\omega_{i}}\right\vert ,
\end{equation}
where $\Delta P_{P,S,I}=P_{P,S,I}^{\mathrm{in}}-P_{P,S,I}^{\mathrm{out}}$.

\begin{figure}
[h]
\begin{center}
\includegraphics[
width=0.65\textwidth
]%
{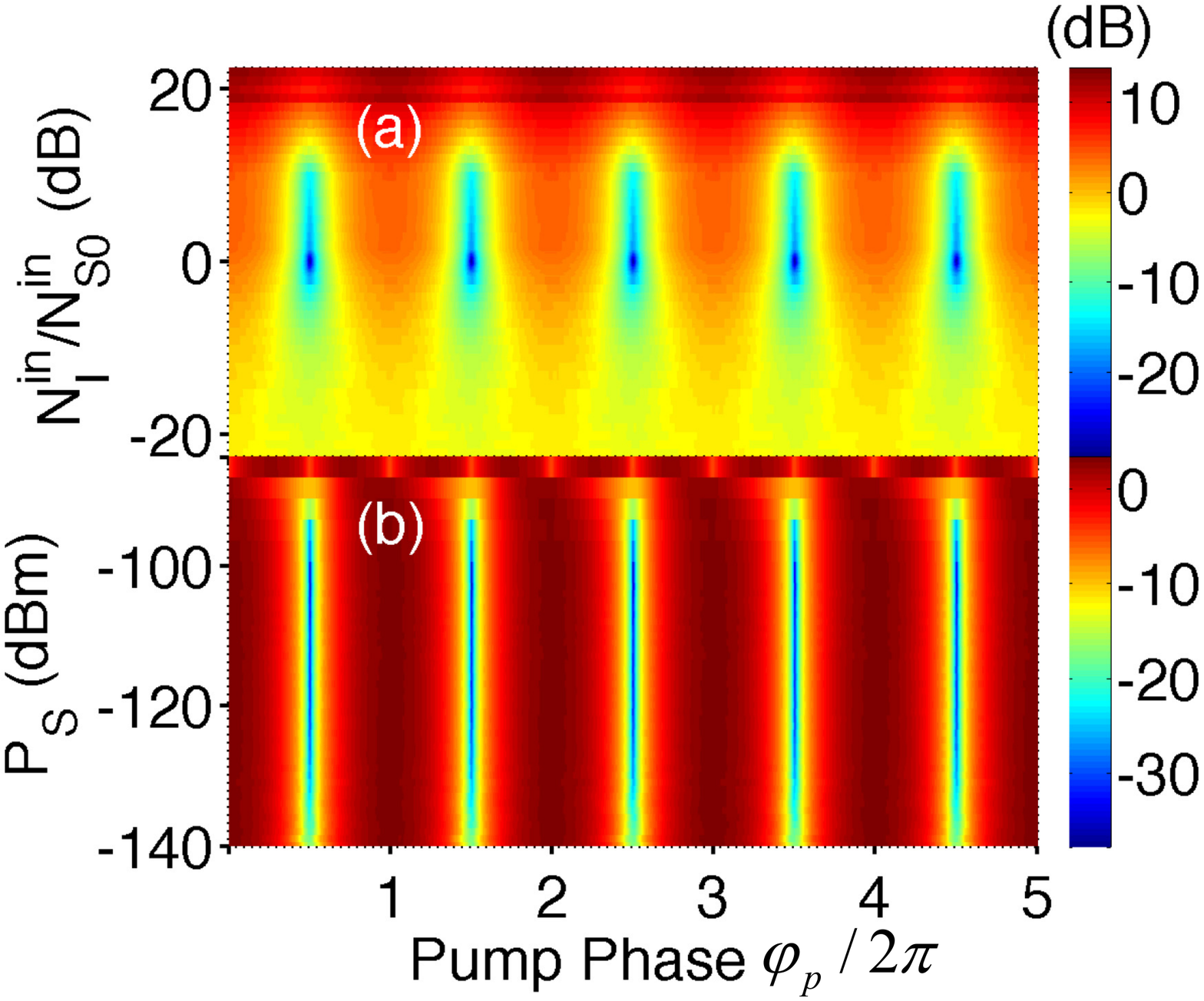}%
\caption{(colored online). Interference fringes measured at the signal port as a function of
the pump phase at the $50/50$ beam-splitting working point. The applied S, I
and P frequencies are identical to the values cited in the main text. In panel (a) the interference is obtained for a varying idler
input photon flux $N_{I}^{\mathrm{in}}$ relative to a constant applied signal
photon flux $N_{S0}^{\mathrm{in}}$ obtained at $P_{S0}$. As expected, the
highest contrast is achieved for the case of $0$ dB, which corresponds to equal I and S photon fluxes. In panel (b) the interference is obtained
by injecting equal input S and I photon fluxes ($N_{I}^{\mathrm{in}}%
=N_{S}^{\mathrm{in}})$ and varying the corresponding input powers by the same
step. In this color plot we show the generated interference as a function of
the corresponding signal input power $P_{S}$. As can be seen in the plot, the
device maintains its high contrast interference up to a maximum input power of
$-100$ dBm above which the device starts to saturate. At very small input
powers below $-130$ dBm the interference modulation amplitude is limited by
the noise floor of the system.}%
\label{varyinput}%
\end{center}
\end{figure}

In the interference measurement shown in Fig. \ref{varyinput} (a) we set
the JPC at the $50$/$50$ beam-splitting working point using the pump tone and
set the input signal power to $P_{S0}$ which corresponds to a signal input
photon flux $N_{S0}^{\mathrm{in}}$ and we vary the idler input photon flux
$N_{I}^{\mathrm{in}}$ relative to $N_{S0}^{\mathrm{in}}$ by varying the input
idler power. As can be seen in the figure, the largest interference contrast
induced by the relative pump phase is obtained at $0$ dB, at which the input
photon fluxes $N_{I}^{\mathrm{in}}$ and $N_{S0}^{\mathrm{in}}$ are equal. In
Fig. \ref{varyinput} (b), we set the JPC at the same
$50$/$50$ beam-splitting working point, but we inject equal signal and idler
photon fluxes and vary the corresponding input powers jointly. This
measurement shows the input power range on the signal port $P_{S}$ over which
the device is capable of maintaining its interference contrast before it
saturates due to nonlinear effects or pump depletion \cite{JRCAreview}. As can
be seen in the figure, the maximum input power which can be handled by the
device is about $-100$ dBm.

\subsection{Electromagnetic Induced Transparency}

It is worthwhile pointing out the strong similarity that exists between
parametric frequency conversion in our Josephson device and the
electromagnetic induced transparency (EIT) effect in optics
\cite{quantumoptics,EITHarris,EITreview,tobiasEIT}. In the EIT effect, the
optical properties (i.e. absorption) of a 3-level atomic medium, as seen
by a weak probe signal that is in resonance with one of the atomic
transitions, are modified by driving the system with a strong coherent Rabi
field in resonance with another atomic transition. In such a 3-level system,
the Rabi drive facilitates a quantum interference between two pathways for the
impinging probe field. As the interference becomes destructive for a certain
drive amplitude, absorption vanishes and the medium becomes transparent.
Likewise, the JPC in conversion mode can be thought as a 3-level
atomic medium whose microwave properties (i.e. total reflection), as seen by
a weak signal, can be changed to perfect absorption (i.e. full conversion)
by inducing destructive interference in the device using a strong Rabi field
that is the pump. Hence, the reflection parameter expressed in Eq.
(\ref{rbb}) is in a way analogous to the complex susceptibility of the medium
in the EIT effect.

\section{Coherent cancellation of output beams}

\begin{figure}
[h]
\begin{center}
\includegraphics[
width=0.65\textwidth
]%
{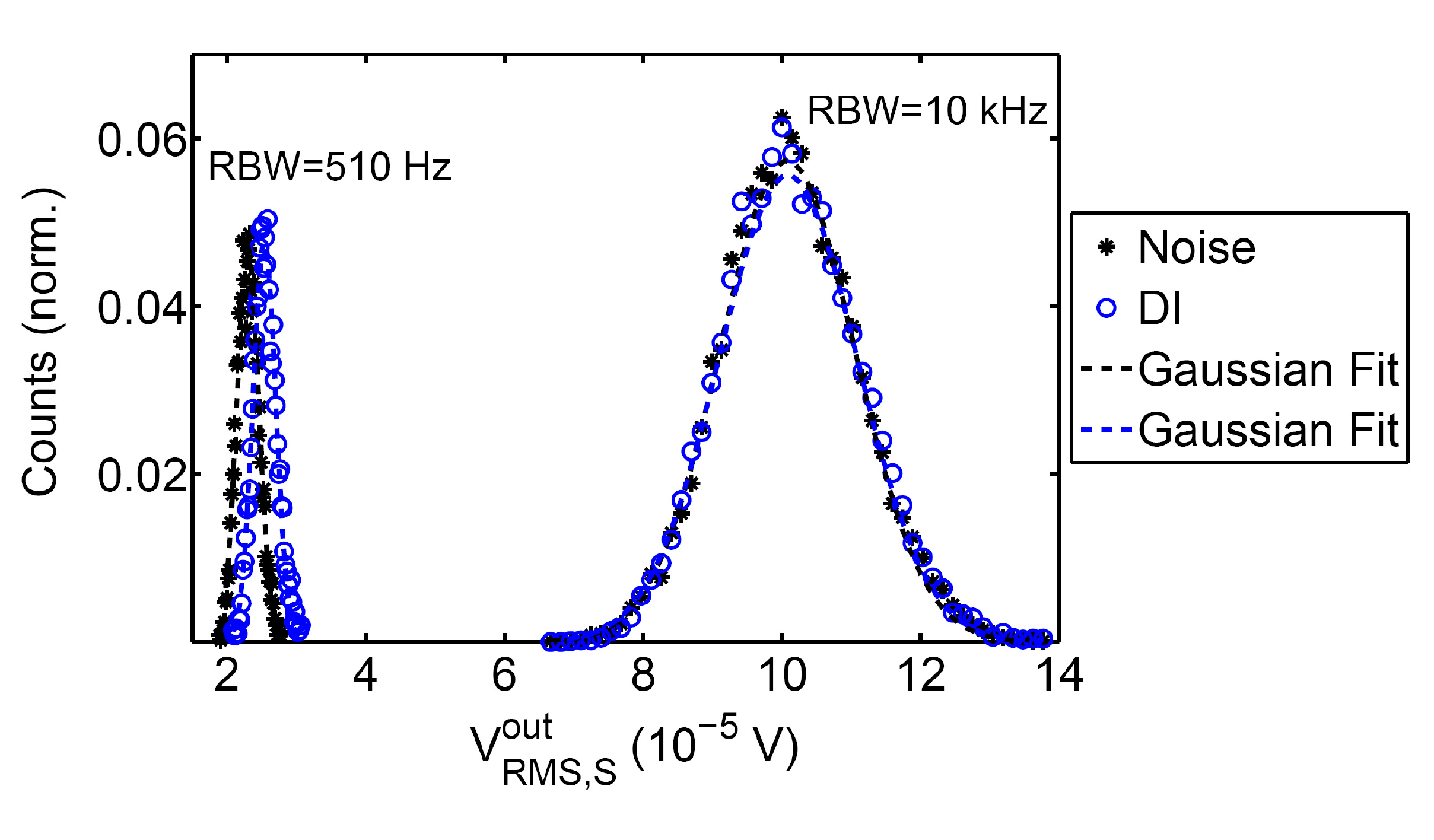}%
\caption{(colored online). Normalized histograms of received rms voltage measured for the system noise (black stars)
and for the destructive interference (DI) experiment (blue circles). The noise
and DI histograms, normalized by the total number of points, are taken using a
spectrum analyzer operated in zero frequency span mode at $f_{S}%
=8.402\operatorname{GHz}$. The different distributions for the noise and DI
plotted on the right and left hand side are measured using two resolution
bandwidths $10\operatorname{kHz}$ and $510\operatorname{Hz}$ respectively. In
the case of the noise measurement, no tone is applied, while in the case of the
DI, the device is operated at the $50/50$ beam-splitting working point and
three continuous wave tones are applied simultaneously: signal, idler and
pump. The parameters of the three applied tones are identical to the values used in the measurement shown in Fig. 5. In this measurement, the pump phase is
manually set to a relative phase which maximizes the DI. The DI working point
is indicated by a green circle in Fig. 5. The histograms on the right
(left) are constructed using $10^{4}$ ($5\cdot10^{3}$) points averaged over
$30$ ($100)$ traces with a sweep time of $10\operatorname{s}$
($1\operatorname{s}$). The traces were measured with equal resolution and
video bandwidths of $10\operatorname{kHz}$ ($510\operatorname{Hz}$). The black
and the blue dashed lines are Gaussian fits to the noise and DI measurements
respectively. }%
\label{ndihist}%
\end{center}
\end{figure}

In order to quantify the amount of coherent cancellation achieved by the
device in a DI experiment at the $50/50$ beam-splitting point, we measure the
received rms voltage distribution of the system noise with no applied tones,
which is set by the output amplification chain, and measure also the received
rms voltage distribution of the signal in a DI condition with three applied
coherent tones as indicated by the green circle in Fig. 5 in the main text. In the DI
case the relative pump phase was manually set using a phase shifter to yield a
maximum destructive interference. In Fig. \ref{ndihist} we plot the measured
noise and DI distributions normalized by the total number of points using
black stars and blue circles respectively. Both distributions are constructed
using averaged time traces of a spectrum analyzer operated in zero frequency
span mode at $f_{S}$. As indicated in Fig. \ref{ndihist}, the data is taken
for two resolution bandwidths $B_{1}=10%
\operatorname{kHz}%
$ and $B_{2}=510%
\operatorname{Hz}%
$ drawn on the right and left sides of the plot. The dashed black curve and
the dashed blue curve are Gaussian fits to the data. The Gaussian means of the
noise distributions shown in Fig. \ref{ndihist}, $\mu_{\mathrm{rms}%
}^{\mathrm{N1,fit}}=10.08\cdot10^{-5}%
\operatorname{V}%
$ for $B_{1}$ and $\mu_{\mathrm{rms}}^{\mathrm{N2,fit}}=2.3\cdot10^{-5}%
\operatorname{V}%
$ for $B_{2}$ agree well with the calculated rms voltage of thermal noise
generated by a $R=50%
\operatorname{\Omega }%
$ load connected to a $50%
\operatorname{\Omega }%
$ transmission line and amplification stages, i.e. HEMT and room temperature
amplifiers, given by $\mu_{\mathrm{rms}}^{\mathrm{N1,calc}}=\sqrt{gk_{B}%
T_{N}B_{1}R}=9.8\cdot10^{-5}%
\operatorname{V}%
$ and $\mu_{\mathrm{rms}}^{\mathrm{N2,calc}}=2.2\cdot10^{-5}%
\operatorname{V}%
$, where $g=81\pm2$ dB is the gain of the signal output chain, $k_{B}$ is
Boltzmann constant and $T_{N}=11\pm2%
\operatorname{K}%
$ is the noise temperature of the output signal line.%

As can be seen in the figure, the noise and DI distributions lie on top of each
other in the first measurement corresponding to $B_{1}$, $\mu_{\mathrm{rms}%
}^{\mathrm{S1,fit}}/\mu_{\mathrm{rms}}^{\mathrm{N1,fit}}=1.002$ , where
$\mu_{\mathrm{rms}}^{\mathrm{S1,fit}}$ is the Gaussian mean of the DI
distribution and they overlap to a large extent in the second measurement
corresponding to $B_{2}$, $\mu_{\mathrm{rms}}^{\mathrm{S2,fit}}/\mu
_{\mathrm{rms}}^{\mathrm{N2,fit}}=1.1$. As can be seen in the figure, the
relative shift between the DI and noise distributions observed in the second
measurement with $B_{2}$, falls well within the standard deviation of the
first measurement with $B_{1}$. Moreover, the larger shift observed in the
second measurement for $B_{2}$ can be attributed to nonidealities of the
measurement, such as the accuracy of the manual setting of the relative pump
phase and the input powers of the three beams as well as the phase stability
of the three generators over time which sets a limit on the amount of
averaging which can be applied in each case.

Furthermore, it is important to note that varying the resolution bandwidth of
the measurement mainly changes the noise floor of system but has no effect on
the received output power originating from the coherent signals applied in the
interference experiment. Hence, by calculating the ratio of the average power
received in the destructive interference experiment measured with $B_{1}$
($B_{2}$) to the average power received in the constructive interference
experiment (not shown in the figure) we get a value of $-25$ dB ($-37.9$ dB)
which corresponds to a coherent cancellation of $99.7\%$ ($99.98\%$) of input
signal and idler photon fluxes.

\section{Comparison with a microwave Mixer}

In this section, we compare our Josephson junction 3-wave beam-splitter/combiner to a
diode based microwave mixer. A microwave mixer has three ports a local
oscillator (LO), a radio frequency (RF) and an intermediate frequency (IF)
ports. An ideal mixer mixes two input signals with different frequencies LO
and IF or LO and RF and generates an output signal at the RF or IF ports which
consists of the sum and difference frequencies of the two inputs. This
frequency conversion property of the device makes it invaluable component for
wireless and cellular communication systems, such as transmitters and receivers
which employ modulation (up-conversion) and demodulation (down-conversion)
schemes. Mixers are also widely used in homodyne and heterodyne microwave
experiments and in measuring devices such as vector network analyzers and
spectrum analyzers. Microwave mixers achieve frequency conversion using
nonlinear elements such as Schottky diodes or field effect transistors (FETs)
which effectively function as nonlinear resistors \cite{Pozar}. However, the nonlinearity
of the diodes and transistors can also generate other high order harmonics and
products of the input frequencies. Hence, a considerable engineering effort is made to
impedance match the three ports and filter undesired products by using
reactive or resistive elements at the different ports.

\begin{figure}
[h]
\begin{center}
\includegraphics[
width=0.65\textwidth
]%
{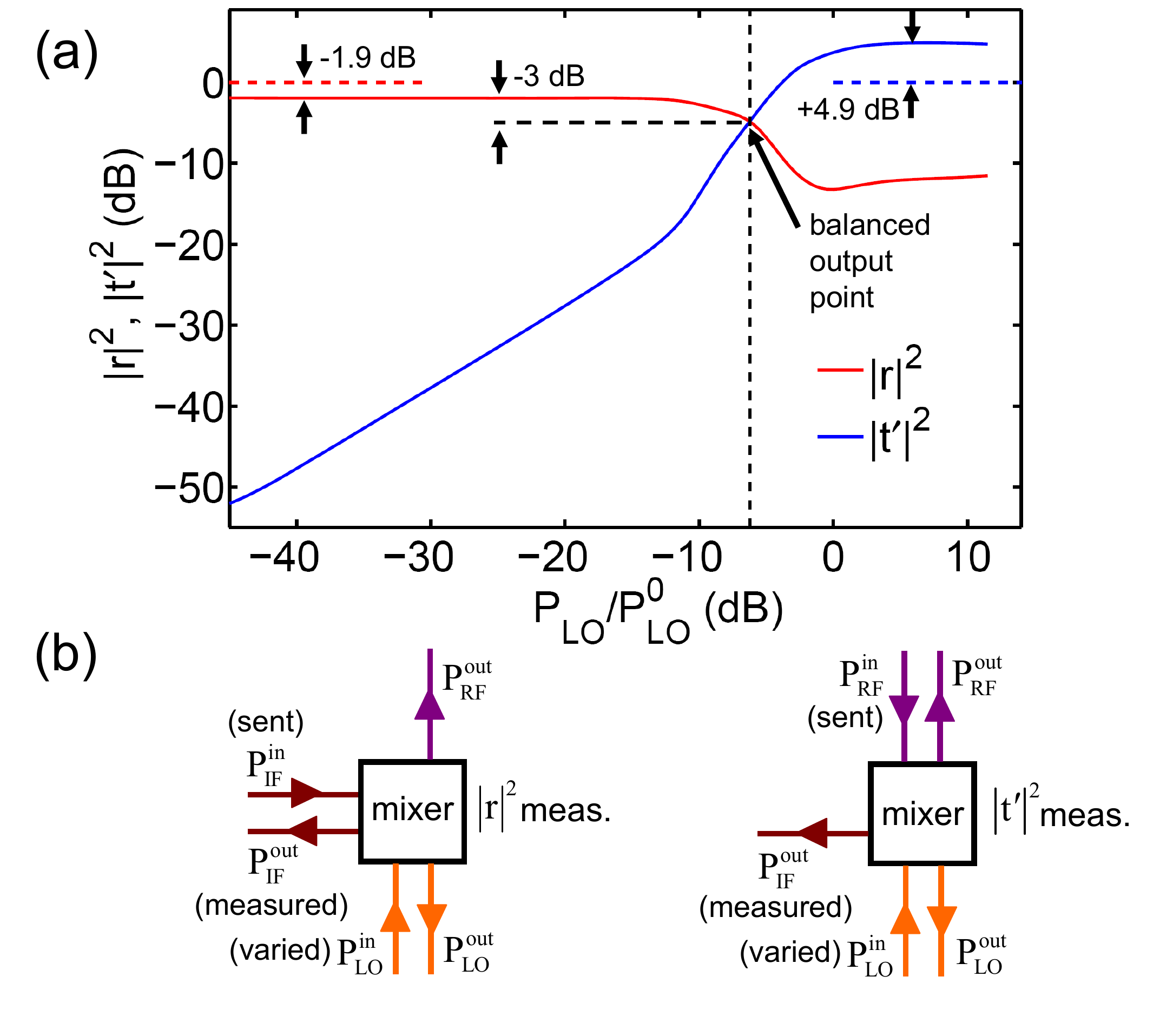}%
\caption{(colored online). Reflection and conversion measurements of a microwave
mixer. (a) The reflection parameter $\left\vert r\right\vert ^{2}$ (red
curve) and the conversion parameter $\left\vert t^{\prime}\right\vert
^{2}$ (blue curve) are plotted in log-log scale as a function of the
normalized local oscillator (LO) power $P_{LO}/P_{LO}^{0}$, where $P_{LO}%
^{0}=6.5$ dBm is the LO power at which $\left\vert r\right\vert ^{2}$ is minimum. Both measurements were taken at the IF port. In the reflection measurement, a coherent tone at $f_{IF}$ was applied to the IF port with input
power $P_{IF}^{\mathrm{in}}$. In the conversion measurement, a coherent
tone at $f_{RF}$ was applied to the RF port with input power $P_{RF}%
^{\mathrm{in}} $, which is set to yield at the balanced output point the same
IF output power as the one measured in reflection for $P_{IF}^{\mathrm{in}}$.
The vertical dashed line indicates the balanced output point of the mixer at
the IF port. (b) A device cartoon showing the signals employed in the
reflection and conversion measurements of the mixer.}%
\label{rtmixer}%
\end{center}
\end{figure}

In the control experiment shown in Fig. \ref{rtmixer}, we measured a commercial off the shelf
triple-balanced mixer by Marki Microwave made of Schottky diodes. The measured
mixer has a wide-bandwidth $2-20$ $%
\operatorname{GHz}%
$ for the RF and LO ports and $1-10$ $%
\operatorname{GHz}%
$ for the IF port. For simplicity, we choose to work with similar frequencies
as in the JPC experiment where the LO, RF and IF play the role of the pump
$f_{LO}=6.285$ $%
\operatorname{GHz}%
$, the idler $f_{RF}=14.687$ $%
\operatorname{GHz}%
$ and the signal $f_{IF}=8.402$ $%
\operatorname{GHz}%
$ respectively.

One of the figures of merit of mixers is conversion loss, which is defined as
$L_{\mathrm{conv}}(dB)=10\log(P_{RF}^{\mathrm{in}}/P_{IF}^{\mathrm{out}})$,
where $P_{RF}^{\mathrm{in}}$ is the RF input power and $P_{IF}^{\mathrm{out}}
$ is the converted power at the IF port. A dual figure of merit can also be
defined for the IF input and the RF output. The main loss mechanisms which
contribute to $L_{\mathrm{conv}}$ are, resistive loads added to the
ports in order to attenuate undesired products, internal losses of the diodes
or FETs, and power conversion to other harmonics or generated products. It is
worthwhile mentioning that the magnitude of conversion loss is also strongly
dependent on the applied LO power.

In Fig. \ref{rtmixer} (a) we plot the reflection ($\left\vert r\right\vert
^{2}=P_{IF}^{\mathrm{out}}/P_{IF}^{\mathrm{in}}$) and the conversion
($\left\vert t^{\prime}\right\vert ^{2}=P_{IF}^{\mathrm{out}}/P_{RF}%
^{\mathrm{in}}$) parameters of the microwave mixer using red and blue curves
respectively as a function of the normalized LO power $P_{LO}/P_{LO}^{0}$,
where $P_{LO}^{0}$ is the LO power at which $\left\vert r\right\vert ^{2}$ is minimum. The reflection and conversion parameters are plotted
in log-log scale in order to cover the several orders of magnitude spanned by
the data. The signals employed in each measurement are sketched in panel (b).
It is important to note that unlike the JPC,\ in a microwave mixer, due to the
reflection and conversion losses, the sum $\left\vert r\right\vert
^{2}+\left\vert t^{\prime}\right\vert ^{2}$ (in linear units) is not equal to
one as can be seen in the figure. The measurement was taken using spectrum
analyzer centered at the IF\ frequency in zero frequency span mode. We also
used a circulator on the IF port of the mixer in order to separate input and
output signals. The input power of the IF tone used in the reflection
measurement is $P_{IF}^{\mathrm{in}}=-43.07$ dBm. The $1.9$ dB loss indicated
in the figure represents the reflection loss of the IF tone when the LO is
turned off ($P_{IF0}^{\mathrm{out}}/P_{IF}^{\mathrm{in}}$). From the
reflection measurement result we find the balanced signal output point of the
device, namely $P_{LO}^{\mathrm{in}}=0.57$ dBm at which the reflected signal
$P_{IF}^{\mathrm{out}}$ drops by $3$ dB relative to $P_{IF0}^{\mathrm{out}}$.
For the conversion measurement, we used an RF tone with input power
$P_{RF}^{\mathrm{in}}=-28.92$ dBm. This input power was set to yield the same
$P_{IF}^{\mathrm{out}}$ ($-47.97$ dBm) measured at the IF port in reflection
at the ``$50/50$ beam-splitting" point. Using the definition of conversion loss
for mixers we find $L_{\mathrm{conv}}=19.05$ dB at the "$50/50$ beam-splitting"
point. As can be seen in the figure, for high LO powers for example at $15$ dBm
the conversion loss improves by about $9.8$ dB and becomes $L_{\mathrm{conv}%
}=9.25$ dB which matches the typical values listed in the data sheet of the mixer.

Furthermore, we can place a lower bound on the inherent power conversion loss
of the mixer which we define as $L_{\mathrm{conv}}^{^{\prime}}%
=L_{\mathrm{conv}}-L_{\mathrm{r}}^{RF}-L_{\mathrm{r}}^{IF}$, where
$L_{\mathrm{conv}}$ is the conversion loss at the optimal LO power at $15$ dBm
and $L_{\mathrm{r}}^{RF}$, $L_{\mathrm{r}}^{IF}$ are the measured reflection
losses at the IF\ ($1.9$ dB) and RF ($1.6$ dB) ports at $f_{IF}$ and $f_{RF}$
respectively. This lower bound $L_{\mathrm{conv}}^{^{\prime}}$ which equals to
$5.7$ dB in the measured mixer shows that it is not possible to model a
microwave mixer made of Schottky diodes using a simple model of a lossless
mixer with the addition of attenuations on the different ports.%

\section{Device and Measurement Details}

The resonators of the JPC are implemented using Nb over a $430$ $%
\operatorname{\mu m}%
$ thick sapphire substrate with a $2$ $%
\operatorname{\mu m}%
$ thick silver ground plane evaporated on the back side, which enhances
thermalization and microwave control. The Nb layer is patterned using a
standard photolithography step and etched using reactive ion etching. The JRM
of the device is incorporated using a standard e-beam lithography process
followed by two angle shadow evaporation of aluminum (with an oxidation step
in between) and lift off. The Josephson junctions of the JRM are nominally
identical with critical current $I_{0}=3\pm0.5%
\operatorname{\mu A}%
$. The Josephson junction area is $5%
\operatorname{\mu m}%
\times1%
\operatorname{\mu m}%
$, while the loop area of the JRM is about $50%
\operatorname{\mu m}%
^{2}$. A large overlap area (partially shown in Fig. 2 (b)) is
established between the Nb part of the resonators and the Al wires of the JRM
which is preceded by plasma cleaning. No observable losses were measured in
our samples due to this interface. In the experiment, the JRM is biased with
half a flux quantum using an external magnetic coil attached to the copper box
housing the device.

The measurements were taken in a dilution fridge at a base temperature of $30
$ m$%
\operatorname{K}%
$. The experimental setup is similar to the one used in Ref. \cite{Jamp}. It
consists of three input lines for the signal, idler and pump, which are
composed of stainless steel semi-rigid coax cables with attenuators at the $4%
\operatorname{K}%
$ and mixing-chamber stages. The input lines of the signal and idler are
connected to the device ports, shown in Fig. 2 (a), through
circulators at base which separate between input and output signals of the
system. The third port of the circulators is connected to two output lines for
the signal and idler. Each output line consists of: (1) two isolators in
series at the mixing-chamber stage with bandwidth $4-12%
\operatorname{GHz}%
$ followed by a low-pass filter with a cutoff frequency at $12$ $%
\operatorname{GHz}%
$ to partially protect the sample from out-of-band noise of the HEMT, (2) a
short semi-rigid coax cable of NbTi which connects between the mixing-chamber
and the $4$ $%
\operatorname{K}%
$ stages, (3) a HEMT amplifier at the $4$ $%
\operatorname{K}%
$ plate with bandwidth $4-8$ $%
\operatorname{GHz}%
$ on the signal side and $4-12$ $%
\operatorname{GHz}%
$ on the idler side, (4) room temperature amplifiers and coax cables. The lack
of commercial cryogenic microwave circulators and isolators and HEMT
amplifiers which work around $15$ $%
\operatorname{GHz}%
$ makes it difficult to measure directly the idler output power of our device.

The interference fringes shown in Figs. 4, 5 and 6 are measured using time traces of a spectrum analyzer in zero
frequency span mode with center frequency $f_{S}$, resolution bandwidth and
video bandwidth of $510$ $%
\operatorname{Hz}%
$. The pump phase was varied as a function of time by offsetting the frequency
$f_{P}=f_{I}-f_{S}$ on the pump generator by a few Hz. The frequency offset
$\Delta f$ was set to be much less than the resolution bandwidth of the
measurement in order to combine the reflected signal at $f_{S}$ and the converted idler at $f_{S}-\Delta f$ in a phase-sensitive manner. Similar
interference fringes can be obtained by alternatively offsetting $f_{S}$ or
$f_{I}$ instead.